\documentclass[preprint,pteplogo]{ptephy}

\preprintnumber{KUNS-2561, YITP-15-42} 

\usepackage{graphicx,color}
\usepackage{multirow}
\usepackage{times}

\newcommand{\fW}{f_{\scriptscriptstyle{W}}}
\newcommand{\fH}{f_{\scriptscriptstyle{H}}}
\newcommand{\Wig}[1]{{#1}_{\scriptscriptstyle{W}}}
\newcommand{\VEV}[1]{\langle{#1}\rangle}

\newcommand{\NTP}{{N_\mathrm{TP}}}
\newcommand{\NMC}{{N_\mathrm{MC}}}
\newcommand{\NIN}{{N_\mathrm{in}}}
\newcommand{\SHW}{{S_\mathrm{HW}}}
\newcommand{\beq}{\begin{eqnarray}}
\newcommand{\eeq}{\end{eqnarray}}
\begin{document}

\title{Entropy production in quantum Yang-Mills mechanics in semi-classical approximation}

\author{\name{Hidekazu Tsukiji}{1,\ast}, \name{Hideaki Iida}{2}, \name{Teiji Kunihiro}{2}, \name{Akira Ohnishi}{1}, and \name{Toru T. Takahashi}{3}}

\address{\affil{1}{Yukawa Institute for Theoretical Physics, Kyoto University,
Kyoto 606-8502, Japan}
\affil{2}{Department of Physics, Faculty of Science, Kyoto University,
Kyoto 606-8502, Japan}
\affil{3}{Gunma National College of Technology, 
Gunma 371-8530, Japan}
\email{tsukiji@yukawa.kyoto-u.ac.jp}}

\begin{abstract}%
We discuss thermalization of isolated quantum systems
by using the Husimi-Wehrl entropy evaluated in the semiclassical treatment.
The Husimi-Wehrl entropy is 
the Wehrl entropy obtained by using the Husimi function
for the phase space distribution.
The time evolution of the Husimi function is given by smearing 
the Wigner function, whose time evolution is obtained 
in the semiclassical approximation.
We show the efficiency and usefulness of this semiclassical treatment
in describing entropy production
of a couple of quantum mechanical systems,
whose classical counter systems are known to be chaotic.
We propose two methods to evaluate
the time evolution of the Husimi-Wehrl entropy,
the test-particle method and the two-step Monte-Carlo method.
We demonstrate the characteristics of the two methods 
by numerical calculations, 
and show that the simultaneous application of the two methods ensures
the reliability of the results of the Husimi-Wehrl entropy at a given time.
\end{abstract}

\subjectindex{A56,A52,D31}

\maketitle

\section{Introduction}

Thermalization process or entropy creation of isolated quantum systems is a long-standing issue, but not
well understood problem.
Relevant  systems include  the early
universe where the transition from a vacuum state to a thermalized
state occurs at the end of cosmic inflation, and the QCD matter created in the initial stage of relativistic heavy-ion
collisions  where thermal matter should be formed in a rather short time.
It is known that both systems are well described in semiclassical approximation, and moreover
a chaotic behavior of the classical limit may play some role in the entropy production.
The present paper is concerned with the entropy production of an isolated quantum system for
which the semiclassical approximation is valid and the classical counter part may show a chaotic behavior.

To describe entropy
in a pure quantum system,
one may of course adopt 
 the von Neumann entropy~\cite{vonNeumann} as quantum mechanical entropy given by
\begin{align}
S_\mathrm{vN}
=& - \mathrm{Tr}\left[\rho \log \rho\right]
\ ,\label{eq:vonNeumann}
\end{align}
where $\rho$ is the density matrix.
For a pure state, however, $\rho$ is idempotent, $\rho^2=\rho$,
implying that the eigen value of $\rho$ is 0 or 1,
and the von Neumann entropy is zero.
Even if we start from a mixed state, the
time evolution described by a unitary operator will never
lead to entropy growth.
On the other hand,
the entropy production in a rarefied gas 
composed of classical or quantum mechanical particles can be well described by 
an analog of the $H$ function of Boltzmann given in terms of the distribution function$f(q, p)$:
\begin{align}
S= - \int \frac{d^Dqd^Dp}{(2\pi\hbar)^D}\,f(q,p)\,\log f(q,p) .
\label{Eq:Boltzmann-H}
\end{align}
It is noteworthy
that a phase-space description is desirable
for making classical-quantum correspondence clear, and
even natural when the semiclassical approximation is valid.
The standard method for such a description is to use 
the celebrated Wigner function~\cite{Wigner},
which is defined as a Wigner transform of the density matrix: 
The Wigner function $\fW(q,p)$ can be regarded as a quasi phase-space
distribution function.
The use of the Wigner function as the phase space distribution function ($f(q,p)=\fW(q,p)$ in Eq.~\eqref{Eq:Boltzmann-H}), 
however,  has essential drawbacks: 
%
First, the Wigner function is, actually, not a genuine distribution function;
$\fW$ can be negative, which
prevents us to calculate the entropy density according to Eq.~\eqref{Eq:Boltzmann-H}.
%
Second, the entropy defined by Eq.~\eqref{Eq:Boltzmann-H} given
in terms of the Wigner function $f_W$ does not grow in time,
because the Wigner transform only gives an equivalent description of the quantum system in terms of, say,
the $q$- or $p$-representation~\cite{Gro46,Moy49,Hillery84,Lee95,Curt14}.
Some coarse graining of the phase space is needed to describe an entropy production.

In a classical chaotic system,
two adjacent points in the phase space 
depart from each other exponentially in time.
If available phase space volume is limited,
the exponentially diffusing classical trajectories have to be folded 
in a complicated manner in the phase space.
After a certain time starting from a localized phase space cell,
a given phase space cell $(2\pi\hbar)^D$ consists of the mixture
of trajectories stemming from the initially
occupied localized cell
and vacant regions not yet visited.
Since we cannot distinguish the phase space points in a cell
due to the uncertainty principle,
it is reasonable to define a phase space distribution
as a smeared or coarse-grained function over the phase space cell.

We adopt the Husimi function $\fH(q, p)$~\cite{Husimi} as such a coarse-grained distribution function,
which is defined as the expectation value 
of the density matrix with respect to a coherent state $\vert z \rangle$.
It is readily shown that $\fH(q, p)$
is {\em semi-positive definite}, $\fH \geq 0$,
and a coarse-grained function of the Wigner function,
as will be shown in a later section.
%
It is shown \cite{TS1985,Takahashi1989} that the Husimi function faithfully describes the characteristic properties of 
the underlying  classical system,
and has been utilized to  identify
the chaotic remnants in quantum systems~\cite{TS1985, Takahashi1989, Sugita03,Sugita-Aiba2002}.
Thus a natural candidate of the quantum mechanical entropy is given by (\ref{Eq:Boltzmann-H})
with $f(q,p)$ being substituted by the Husimi function $f_H(q, p)$.
This entropy was introduced by Wehrl~\cite{Wehrl} and may be called the Wehrl entropy,
although he himself called it the classical entropy and failed in identifying the distribution function
$\fH(q, p)$ with the Husimi function:
Such an identification was made later ~\cite{Anderson1993}. 
We refer to the Wehrl entropy obtained by using the Husimi function
as the {\em Husimi-Wehrl (HW) entropy}~\cite{KMOS},
\begin{align}
S_\mathrm{HW}= - \int \frac{d^Dqd^Dp}{(2\pi\hbar)^D}\,\fH(q,p)\,\log \fH(q,p)\ .
\label{Eq:HWE}
\end{align}
It is worth mentioning that the HW entropy can be a good measure for a 
quantum entanglement of a system including quantum optical systems~\cite{MZ2004,AA2007}.
For a one-dimensional case, 
there is a minimum of $S_\mathrm{HW}=1$~\cite{Lieb,Wehrl1979},
in contrast to the von Neumann entropy, 
which takes $S_\mathrm{vN}=0$ in the ground state.
It is also shown that the HW entropy takes a value close to the von Neumann
entropy at high temperature, and its growth rate coincides
with the Kolmogorov-Sina\"{i} entropy
for the one-dimensional  inverted harmonic oscillator~\cite{KMOS}.

A direct evaluation of the HW entropy for a quantum system is a kind of challenge even for 
the system with a few degrees of freedom because it involves a large-dimensional
integral over the phase space even apart from the cumbersome calculation of the
logarithm with precision.
Nevertheless the HW entropy and its time evolution have been calculated for some quantum 
systems~\cite{OtherHWE,TsaiMuller}. The equation of motion (EOM) of the Husimi function is
given in \cite{Takahashi1989}, which contains a term of the order $\hbar$, and thus has a more complicated form
than that of the Wigner function even in the semiclassical approximation; see below.
To solve the complicated EOM of the Husimi function,
a test-particle method was proposed
by Tsai and Muller~\cite{TsaiMuller},
where the evolution of the test particles are determined
to reproduce some of the moments.

%
As already mentioned, the semiclassical approximation
is suitable to reveal the effect of the chaotic nature 
of the classical counter part. 
It is noteworthy  that the time evolution of the Wigner function 
in the semiclassical approximation
where the $\mathcal{O}(\hbar^2)$ terms are ignored
is readily obtained by solving
the classical Hamilton equation;
quantum mechanical information such as the uncertainty relation
is encoded in the initial Wigner function,
provided that it is given as the Wigner transform 
of the quantum density matrix.


The time evolution of the Husimi function
is given by smearing the time-evolved Wigner function
obtained in the semiclassical approximation.
This is the method we adopt in this article.
We shall show its efficiency and usefulness
in describing entropy production
using a couple of quantum mechanical systems
whose respective classical counter systems are known to be chaotic.
We propose two methods to evaluate
the time evolution of the Husimi-Wehrl entropy.
One is an adaptation of the usual test-particle method without recourse to
the moments of the distribution function.
The other is a sequential application of Monte-Carlo integration,
which we call the two-step Monte-Carlo method.
We shall demonstrate the characteristics of the two methods 
by numerical calculations, 
and show that the simultaneous application of the two methods ensures the
reliability of the results of the HW-entropy's time evolution.
It should be noted that these two methods are, in principle, applicable
to systems with large degrees of freedom such as quantum field theories.

The paper is organized as follows.
In Sec.~\ref{sec:review},
we summarize some basic ingredients of the Wigner and Husimi functions
together with the HW entropy. 
In Sec.~\ref{sec:methods},
we introduce the two numerical methods to evaluate the HW entropy 
in an efficient way.
In Sec.~\ref{sec:results},
the quantum mechanical models are introduced
and numerical results of the Husimi-Wehrl entropy are shown.
The final section is devoted to a brief summary and concluding remarks.

\section{Wigner function, Husimi function, and Husimi-Wehrl entropy}
\label{sec:review}

In this section, we briefly review quantum mechanical 
phase space distribution functions,
Wigner~\cite{Wigner} and Husimi~\cite{Husimi} functions,
and the phase space expression of the entropy, Husimi-Wehrl entropy~\cite{Wehrl}.
While
we introduce Wigner and Husimi functions
in one-dimensional quantum mechanics
in Subsec.~\ref{subsec:WignerHusimi}
and \ref{subsec:SemiClassical}, 
extension to multi-dimensional cases is straightforward.

\subsection{Wigner and Husimi functions}
\label{subsec:WignerHusimi}
The Wigner function~\cite{Wigner} is defined as a Wigner transform 
of the density matrix
\begin{align}
\fW(q,p,t)
=& \Wig{\rho}(q,p,t)
\nonumber\\
\equiv& \int d\eta\,
e^{-ip\eta/\hbar}
\VEV{q+\frac{\eta}{2}\mid\rho(t)\mid q-\frac{\eta}{2}}\ .
\label{Eq:Wigner}
\end{align}
While the Wigner function $\fW(q,p)$ 
can be regarded as a quasi phase space distribution function and
provides intuitive picture of the phase space dynamics,
it is not semi-positive definite 
and hence we cannot regard
$\fW(q,p)$ as the phase space probability density.

In order to overcome
the above drawbacks of the Wigner function,
Husimi introduced a Gaussian smeared Wigner function~\cite{Husimi},
known as the Husimi function,
\begin{align}
\fH(q,p)
=&\int \frac{dq' dp'}{\pi\hbar} 
e^{-\Delta(q-q')^2/\hbar-(p-p')^2/\Delta\hbar}\,
\fW(q,p)
\ ,
\label{Eq:Husimi}
\end{align}
where $\Delta$ is an arbitrary width parameter
that gives the smearing manner in the phase space.

The Husimi function is defined also as the expectation value 
of the density matrix with respect to a coherent state $\vert z \rangle$:
\beq
\fH(q, p)=\langle z \vert \rho \vert z \rangle,\quad z =(\Delta q+ip)/\sqrt{2\hbar \Delta},
\label{Eq:HusimiZ}
\eeq
for a one-dimensional case with $\Delta$ being an arbitrary constant.
Here the coherent state is given by 
\beq
\vert{z}\rangle=
e^{z {a}^{\dag}-z^{\ast} {a}}\vert 0\rangle, \quad 
{a}=(\Delta \hat{q}+i\hat{p})/\sqrt{2\hbar \Delta},
\eeq
where $\vert 0\rangle $ is the ground state; $\hat{a}\vert 0\rangle=0.$ 
It is readily shown that $\fH(q, p)$
is {\em semi-positive definite}, $\fH \geq 0$
by using Eq.~\eqref{Eq:HusimiZ}; 
$\fH=\left|\langle{z}\vert\psi\rangle\right|^2 \geq 0$
for a pure state $\vert{\psi}\rangle$,
and $\fH=\sum_i w_i \left|\langle{z}\vert\psi_i\rangle\right|^2 \geq 0$
for a mixed state specified by the density matrix
$\rho = \sum_i w_i \vert\psi_i\rangle\langle\psi_i\vert (w_i \geq 0)$.

The Husimi function $\fH(q,p)$ serves
as the probability density to observe the phase space
variables $(q,p)$ under a minimum wave packet $|z\rangle$,
and is now semi-positive definite, $\fH \geq 0$.
Compared with the Wigner function, 
the Husimi function is smooth and 
the peak of the Husimi function often appears around the expectation value
of the position and momentum~\cite{Takahashi1989, Sugita}.

\subsection{Time evolution in semiclassical approximation}
\label{subsec:SemiClassical}

The equation of motion (EOM) for the Wigner function $\fW$ is obtained 
from the Wigner transform of the von Neumann equation for the density matrix,
$\partial\rho/\partial t=[H,\rho]/i\hbar$.
By applying the Wigner transform of the operator product,
$\Wig{(AB)}=\Wig{A}\exp(i\hbar(\overleftarrow{\nabla}_q \overrightarrow{\nabla}_p
-\overleftarrow{\nabla}_p \overrightarrow{\nabla}_q)/2)\Wig{B}$~\cite{Gro46,Moy49,Curt14},
commutators are replaced by  Poisson brackets as
$\Wig{[A,B]}/i\hbar=\{A,B\}_\mathrm{PB}+\mathcal{O}(\hbar^2)$.
Thus the EOM for $\fW$ is given in terms of the Wigner transform $\Wig{H}$ of the
Hamiltonian $H$ as
\begin{align}
\frac{\partial\fW}{\partial t}
=&\left\{ \Wig{H}, \fW \right\}_\mathrm{PB}
+\mathcal{O}(\hbar^2)
\ .\label{Eq:EOMfW}
\end{align}
The Wigner transform $\Wig{H}$ of a Hamiltonian with the form of $H=p^2/2m + U(q)$
 does not change its form.
We note that the $\mathcal{O}(\hbar^2)$ term in (\ref{Eq:EOMfW}) is proportional to
the third derivative of $\Wig{H}$ or $U$. Thus  the EOM (\ref{Eq:EOMfW}) without
the $\mathcal{O}(\hbar^2)$ term turns out to be exact for some simple models such as
a (an inverted) harmonic oscillator.

The semiclassical EOM for $\fW$ is given by retaining the terms up to $\mathcal{O}(\hbar)$
in Eq.~\eqref{Eq:EOMfW}, which reads
\begin{align}
\frac{\partial\fW}{\partial t}
+\frac{\partial \Wig{H}}{\partial p}\,\frac{\partial \fW}{\partial q}
-\frac{\partial \Wig{H}}{\partial q}\,\frac{\partial \fW}{\partial p}
=0
\ .
\label{Eq:EOM}
\end{align}
We remark that the semiclassical EOM is exact for the linear systems mentioned above.
Equation \eqref{Eq:EOM} asserts that $\fW$ is constant along the classical
trajectory:
Let us see this. Let $(q(t; \bar{q}),\, p(t; \bar{p}))$ is 
a solution of 
the classical EOM, i.e., Hamilton's equation;
\begin{align}
\frac{dq}{dt}=&\frac{\partial \Wig{H}}{\partial p}
\ ,\quad
\frac{dp}{dt}=-\frac{\partial \Wig{H}}{\partial q}
\ ,\label{Eq:Canonical}
\end{align}
with an initial condition $(q(0)=\bar{q}, p(0)=\bar{p})$.
Then we have for $\fW(q(t; \bar{q}),\, p(t; \bar{p}),\, t)$,
\begin{align}
\frac{D\fW}{Dt}\equiv \frac{\partial\fW}{\partial t}
+\frac{dq}{dt}\frac{\partial\fW}{\partial q}
+\frac{dp}{dt}\frac{\partial\fW}{\partial p}
=0\ ,
\label{Eq:Wconst}
\end{align}
which implies that $\fW$ is time-independent;\,
$\fW(q(t; \bar{q}),\, p(t; \bar{p}),\, t)=\fW(\bar{q},\, \bar{p},\, 0)$.
Accordingly we have
\begin{align}
\fW(q,\, p,\, t)=\fW(q(-t; q),\, p(-t; p),\, 0).
\label{Eq:Wconst-1}
\end{align}
Thus
we can 
obtain the semiclassical time evolution of the Wigner function
by solving the classical equation of motion. 
Note that the quantum mechanical 
effects are taken into account through the distribution of the initial value in the phase space  
encoded in the Wigner function $\fW(q,\, p,\, 0)$ constructed
from the initial density matrix.

It is worth mentioning that the exact analytical solution of the time
evolution of $\fW$ for some linear systems including
a (stable) harmonic oscillator potential~\cite{Gro46,KMOS},
an inverted (unstable) harmonic oscillator potential~\cite{KMOS,OtherIHO}
and an external potential~\cite{KMOS} can be obtained.
Then  even the analytic form of the Husimi function $\fH(q,\,p,\,t)$ for these systems
are readily obtained \cite{KMOS} by the Gaussian smearing of $\fW(q,p,t)$, 
which is easy to perform analytically.

We note here that one may obtain the time evolution of the Husimi function $\fH(q,\, p,\, t)$ 
by solving the EOM for  $\fH(q,\, p,\, t)$,
which involves terms proportional to $\hbar$, and thus 
has a more complicated structure than that for $\fW(q,\,p,\, t)$ 
even in the semiclassical approximation~\cite{Takahashi1989}.
%
If one sticks to solve the EOM for $\fH$ directly, some numerical method would be necessary.  
A test-particle method is adopted as such a numerical method by Tsai and Muller~\cite{TsaiMuller},
where the time evolution of test particles are determined
so as to reproduce some of moments. We remark that there are some ambiguities in such an approach
inherent in the moment method.

%
%
%
%

In this work, we do not adopt this direct method for obtaining
the time evolution of the Husimi function $\fH(q,\,p,\,t)$.
We take advantage of the fact that the EOM of the Wigner function $\fW(q,\,p,\, t)$ 
in this regime is given simply by solving the classical EOM, and obtain 
$\fH(q,\,p,t)$ by
the Gaussian smearing of thus obtained $\fW(q,\,p,\,t)$.
This strategy should be workable and natural when the semiclassical approximation 
is meaningful.
The remaining task that we have to do for obtaining the Husimi function is 
 just the multi-dimensional integrations over the phase space with the Gaussian kernel for the smearing,
which should be feasible by  standard methods such as the Monte-Carlo integration.

\subsection{Husimi-Wehrl entropy}
\label{subsec:Wehrl}

%
Since the Wigner function $\fW$ is merely the Weyl transform of the density matrix, 
any observable is calculable in terms of $\fW$ in principle, and it is also the case with the 
Husimi function $\fH$. A drawback of the $\fW$ is that it can have negative values, and hence
is not suitable for the calculation of entropy.
As is mentioned in Introduction and the previous subsection,
the Husimi function is, in contrast, a semi-positive definite
{\em coarse-grained} phase space distribution function
smeared by a minimum wave packet, and hence
a good candidate for
the phase space distribution $f(q,p)$ to evaluate the entropy of 
a quantum system, 
as the $H$ function of Boltzmann in the classical system,
Eq.~\protect\eqref{Eq:Boltzmann-H},
or equivalently the Husimi-Wehrl entropy given in Eq.~\eqref{Eq:HWE}~\cite{Wehrl}.
%

An explicit form of the HW entropy in terms of the Wigner function
is given by
substituting the $D$-dimensional extension of Eq.~\eqref{Eq:Husimi} 
into Eq.~\eqref{Eq:HWE},
\begin{align}
S_\mathrm{HW}(t)
=&-
\int \frac{d^D q d^D p}{(2\pi \hbar)^D} 
\int \frac{d^D q' d^D p'}{(\pi\hbar)^D} 
e^{-\Delta(q-q')^2/\hbar-(p-p')^2/\Delta\hbar}\,
\fW(q',p',t)\nonumber\\
&\times \log \left[\int \frac{d^D q'' d^D p''}{(\pi\hbar)^D} 
e^{-\Delta(q-q'')^2/\hbar-(p-p'')^2/\Delta\hbar}\,
\fW(q'',p'',t)\right]
\ .\label{Eq:HWE2}
\end{align}

One may now recognize some difficulty of the  numerical evaluation of
 the HW entropy: It involves repeated numerical integrations  over the multi-dimensional phase space,
and in particular one of them appears as an argument of logarithm, which
 turns out to be quite problematic in the Monte-Carlo integration.

\section{Numerical methods to analyze the semiclassical time evolution of Husimi-Wehrl entropy}
\label{sec:methods}

Here, two numerical methods are introduced to
calculate the time dependence of the HW entropy 
as given by the Gaussian smearing of the Wigner function obtained
in the semiclassical approximation. Both methods are based on
an adaptation of
the Monte-Carlo integration over the phase-space.
We call the two methods the test-particle (TP)
and two-step Monte-Carlo (tsMC) methods, respectively.
In this section, we deal with the $D$-dimensional system described by the
Hamiltonian $H=H(q,\,p)$, where $q$ and $p$ denote the $D$-dimensional vector, respectively, i.e.,
$q=(q_1,\,q_2,\dots\, ,q_D)$ and $p=(p_1,\,p_2,\dots\, ,p_D).$ 

\subsection{Test-particle method}

In the test-particle method~\cite{TPplasma,Wong1982,TPtext,GuideBUU},
the Wigner function is represented as a sum of the delta functions,
\begin{align}
\fW(q,\,p,\,t)=&\frac{(2\pi\hbar)^D}{N_\mathrm{TP}}\sum_{i=1}^{N_\mathrm{TP}}
\delta^D(q-q_i(t))\,\delta^D(p-p_i(t))
\ ,\label{Eq:WignerTP}
\end{align}
with the initial function
\[
\fW(q,p,0)=\frac{(2\pi\hbar)^D}{N_\mathrm{TP}}\sum_{i=1}^{N_\mathrm{TP}}\delta^D(q-q_i(0))\,\delta^D(p-p_i(0)),
\]
where $N_{\rm TP}$ is the total number of the test particles,
and their coordinates are given by $(q_i(t),\, p_i(t))$.
The initial distribution of the test particles 
$(q_i(0),\,p_i(0))$\, $(i=1,\,2,\dots,\,D)$
is chosen so as to well sample that of $\fW(q,\, p,\,0)$:
Hence $\NTP$ is called the sampling number.
The time evolution of the coordinates
 $(q_i(t),p_i(t))$ 
is determined by the EOM for $\fW(q,\,p,\,t)$, which is reduced to
 the canonical equation of motion,
\begin{align}
\frac{dq_i}{dt}=&\frac{\partial \Wig{H}}{\partial p_i}
\ ,\quad
\frac{dp_i}{dt}=-\frac{\partial \Wig{H}}{\partial q_i}
\ ,\label{Eq:Canonical-D}
\end{align}
in the semiclassical approximation.

For the test-particle representation 
of the Wigner function Eq.~\eqref{Eq:WignerTP},
the Husimi function is readily expressed as 
\begin{align}
\fH(q,p,t)=&\frac{2^D}{\NTP} \sum_{i=1}^{\NTP} 
e^{-\Delta (q-q_i(t))^2/\hbar-(p-p_i(t))^2/\Delta\hbar}
\ .\label{Eq:HusimiTP}
\end{align}
It is noteworthy that the 
Husimi function here is a smooth function
in contrast to the corresponding Wigner function in Eq.~\eqref{Eq:WignerTP}.

Inserting the Wigner function \eqref{Eq:WignerTP} into Eq.~\eqref{Eq:HWE2}, 
the HW entropy in the test-particle method is 
given as,
\begin{align}
S_\mathrm{HW}^\mathrm{(TP)}
=&- 
\frac{1}{\NTP}
\sum_{i=1}^\NTP
\int \frac{d^Dqd^Dp}{(\pi\hbar)^D}\,
e^{-\Delta (q-q_i(t))^2/\hbar-(p-p_i(t))^2/\Delta\hbar}
\log \fH(q,p,t).
\end{align}
Now note that the integral over $(q,p)_i$ for each $i$ has a support 
only around the positions of 
the test particles $(q_i(t),\,p_i(t))$
due to the Gaussian function,
and then we can effectively perform the Monte-Carlo integration
as follows;
By generating a set of random numbers $(Q,P)_i$
with standard deviations of $\sqrt{\hbar/2\Delta}$ and $\sqrt{\hbar\Delta/2}$,
Monte-Carlo sampling point $(q,p)_i$ for each $i$ is obtained as
$(q,p)_i=(Q,P)_i+(q_i,p_i)$. Thus we reach  the formula to be used 
in the actual evaluation of the HW entropy in the test-particle method:
\begin{align}
S_\mathrm{HW}^\mathrm{(TP)}
\simeq&-
\frac{1}{\NMC\NTP}\sum_{k=1}^\NMC\sum_{i=1}^\NTP
\log\left[
\frac{2^D}{\NTP} \sum_{j=1}^{\NTP} 
e^{-\Delta (Q_k+q_i(t)-q_j(t))^2/\hbar-(P_k+p_i(t)-p_j(t))^2/\Delta\hbar}
\right]
\ ,
\label{Eq:HWEtp}
\end{align}
where the amount of the sample number of $(Q,P)_i$ is denoted by $\NMC$.

\subsection{Two-step Monte-Carlo method}

The second method is a direct Monte-Carlo evaluation of the multi-dimensional integrals.
We rewrite Eq.~\eqref{Eq:HWE2} as
\begin{align}
S_\mathrm{HW}^\mathrm{(tsMC)}
=& - 
\int \frac{d^DQd^DP}{(\pi\hbar)^D}\,e^{-\Delta Q^2/\hbar-P^2/\Delta\hbar}
\int \frac{d^Dqd^Dp}{(2\pi\hbar)^D}\,\fW(q,p,t)
\nonumber\\
&\times\log\left[
\int \frac{d^DQ'd^DP'}{(\pi\hbar)^D}\,e^{-\Delta (Q')^2/\hbar-(P')^2/\Delta\hbar}
\,\fW(q+Q+Q',p+P+P',t)
\right]
\nonumber\\
\simeq & - 
\frac{1}{N_\mathrm{out}} \sum_{k=1}^{N_\mathrm{out}}
\log\left[
\frac{1}{N_\mathrm{in}} \sum_{l=1}^{N_\mathrm{in}}
\,\fW(q_k+Q_k+Q'_l,p_k+P_k+P'_l,t)
\right]
\nonumber\\
=&
-\left\langle
\log\left\langle{\fW(q+Q+Q',p+P+P',t)}\right\rangle_{Q'P'}
\right\rangle_{QPqp}
\ ,
\label{Eq:HWEtsMC}
\end{align}
where $(Q_k,P_k)$ and $(Q'_l,P'_l)$ are Gaussian random numbers
for the Monte-Carlo (MC) integration to compute Husimi function $\fH(q,p)$.
For the $(q,p)$-integration, we generate MC samples $(q',p')$ at $t=0$ according to
the initial distribution, and obtain the corresponding phase space sample points
$(q(q',p',t),p(q',p',t))$ at $t$
by solving the canonical equation of motion.
Under the semiclassical approximation,
$\fW$ is constant and the Jacobian is unity along the classical trajectory,
$J(q(t),p(t)/q'(0),p'(0))=1$.
Then we can replace the integral over $(q,p)$ in the first line
of Eq.~\eqref{Eq:HWEtsMC}
with the integral at $t=0$ by using the initial distribution
and the Liouville theorem as,
\begin{align}
&\int \frac{d^Dqd^Dp}{(2\pi\hbar)^D}
\fW(q,p,t) g(q,p)
\nonumber\\
=&\int \frac{d^Dq'd^Dp'}{(2\pi\hbar)^D}
\fW(q',p',0)\, g(q(q',p',t),p(q',p',t))
\ ,
\end{align}
where $(q',p')$ are the phase space coordinates at $t=0$,
and $(q(q',p',t),p(q',p',t))$ are those at $t$ evolved from $(q',p')$.

The Wigner function at $t$ in the log in Eq.~\eqref{Eq:HWEtsMC}
can be 
obtained by the trace back of the trajectory from $t$ to $t=0$
as shown in  Eq.(\ref{Eq:Wconst-1}).
Equation~\eqref{Eq:HWEtsMC} contains 
an MC integral of a function obtained by an MC integral;
we first generate $(q',p')$ at $t=0$ according to the distribution
$f_W(q,p,0)$ and $(Q,P)$ as Gaussian random numbers,
and then perform the MC integral in the log by generating MC samples $(Q',P')$.
We call this procedure {\em two-step Monte-Carlo} (tsMC).

In the following sections, we 
show the characteristic properties of the two methods and
demonstrate numerically how they work
using two-dimensional
quantum-mechanical systems.

\section{Numerical calculation of Husimi-Wehrl entropy in quantum Yang-Mills model}
\label{sec:results}
In this section, we show the numerical results of the HW entropy
in ``quantum Yang-Mills system'' \cite{qYMref},
obtained by the two distinct methods, TP and tsMC methods.

\subsection{Model Hamiltonian and setup of initial condition}

The  Hamiltonian of the system 
is given by
\begin{align}
H=\frac{1}{2m}(p_1^2+p_2^2)+\frac{1}{2}q_1^2q_2^2.
\label{qYMHamiltonian}
\end{align}
We have restricted ourselves to the two-dimensional case here.
The name, ``quantum Yang-Mills (qYM)'', is originated from the 
fact that the spatially uniform Yang-Mills system 
is reduced to a $(0+1)$-dimensional system, i.e., 
a quantum mechanical system, and its Hamiltonian 
is just given by Eq.~\eqref{qYMHamiltonian}. 

We adopt the initial condition
given by a minimal wave packet centered at $(q_1, q_2, p_1, p_2)=(0,0,10,10)$,
\begin{align}
f_{\rm W}(p_1,p_2,q_1,q_2,t=0)
=4e^{-[q^2_1+q^2_2+(p_1-10)^2+(p_2-10)^2]/\hbar}.
\label{Eq:initcond}
\end{align}
This initial condition is also adopted in Ref.~\cite{TsaiMuller}.

In the following, we show numerical results calculated by using the TP
and tsMC methods.
We show the results in the unit with $m=1$ and
$\hbar=1$, and take $\Delta=1$ for the wave packet width.
In the case of $\Delta \neq 1$, the smearing Gaussian is not symmetric in p and q directions. But the results do not change qualitatively.
We have confirmed that the results with $\Delta=0.1$ and $10$ are
qualitatively
the same as those with $\Delta=1$.

\subsection{Numerical results with TP method}
\label{subsec:qYM-TP}

First,
we show the numerical results of the HW entropy in the qYM system
calculated in the TP method using Eq.~\eqref{Eq:HWEtp}.

\begin{figure}[t]
\begin{center}
\includegraphics[width=8cm]{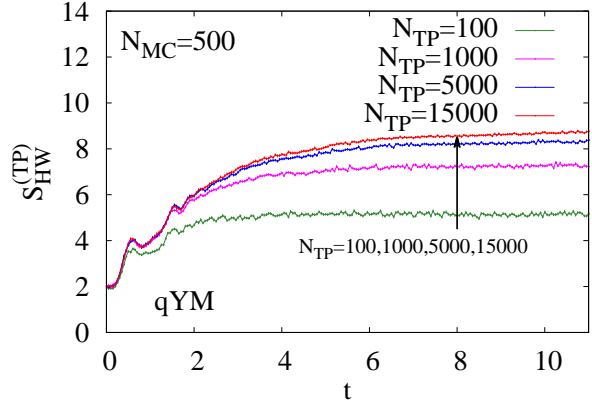}
\caption{Time dependence of the HW entropy by using TP method in qYM,
with $N_{\rm TP}=100, 1000, 5000$ and $15000$, and $N_{\rm MC}=500$.
The arrow shows how the calculated HW entropy changes 
as $N_{\rm TP}$ increases.}
\label{TP_QYMm}
\end{center}
\end{figure}

Figure \ref{TP_QYMm} shows the time evolution of the HW entropy
calculated in the TP method with the following test-particle numbers,
$N_{\rm TP}=100, 1000, 5000$ and $15000$.
The MC sample number is taken to be $N_{\rm MC}=500$.
The statistical errors are estimated for $\NMC$ samples from a standard deviation.
We note that the calculated HW entropy at each $t$ tends to increase
along with increasing $N_{\rm TP}$,
which is an artifact due to the small number of the test particles
$N_{\rm TP}$
and discussed later.
Apart from tiny fluctuations,
all the calculation show that the HW entropy first increases
in time
with a small oscillatory behavior being accompanied;
its local maxima are seen around $t\simeq 0.5$ and $1.7$.
We note that a similar behavior is also seen in Ref.~\cite{TsaiMuller}.
%


Entropy evaluated by the TP method has a (unphysical) maximum depending
on $N_{\rm TP}$, which causes apparent saturation at large $t$ in Fig.~\ref{TP_QYMm}.
In fact, when the system is chaotic and the phase space volume is very large, 
all the test particles will be so separated 
from each other in the phase space at later time
that only the $i=j$ terms in Eq.~\eqref{Eq:HWEtp} will remain.
In this limiting case, 
the HW entropy as given in \eqref{Eq:HWEtp} is evaluated as follows;
\begin{align}
S_\mathrm{HW}^\mathrm{(TP)}
\to&- 
\left\langle
\left[
\log\left(\frac{2^D}{\NTP}\right)
-\Delta Q^2/\hbar-P^2/\Delta\hbar
\right]
\right\rangle_{QP}
\nonumber\\
&=D-D\log{2}+\log{\NTP}
\ ,
\label{eq:limiting}
\end{align}
which gives the inevitable upper limit of $S_{\rm HW}^{({\rm TP})}$.
In Appendix A, we
examine the HW entropy of an inverted harmonic oscillator,
for which $S_{\rm HW}$ can be calculated analytically and is
found to increase permanently.
At later times, $\SHW$ is underestimated with small $\NTP$ values
because of the upper limit discussed above.
By comparison,
$\SHW$ at early times is calculated precisely in the TP method,
as long as $\NTP$ is large enough for $\SHW$ to converge.

From the above argument,
$\SHW(t)$ would be obtained reliably
as an extrapolated value in the limit of $\NTP \to \infty$.
The extrapolation should be made in the $\NTP$ range,
where the limiting value is larger than the HW entropy to be obtained.
The
limiting values are
$S_\mathrm{HW}^\mathrm{(TP)}=5.2, 7.5, 9.1$ and $10.2$
for $\NTP=100, 1000, 5000$ and $15000$, respectively.
The large-t values found in Fig.~\ref{TP_QYMm} are close
to these limiting values for smaller $\NTP$, i.e.,
$\NTP = 100$ and $1000$.
Thus we see that 
the saturation behavior seen for smaller values of $N_\mathrm{TP}$ 
may be an artifact of the TP method.
In contrast
,
the large-t values for $N_\mathrm{TP} = 5000$ and $15000$ 
in Fig.~\ref{TP_QYMm}
are well below the limiting values 
($9.1$ and $10.2$),
found free from the above mentioned artifact, 
and can be used to obtain the extrapolated value at $\NTP\to\infty$,
as discussed later in Subsec.~\ref{subsec:CompTP}.
Thus we conclude that
the entropy production of the ``quantum Yang-Mills'' system 
can be well described with the use of HW entropy as calculated 
with the TP method
with sufficiently large number of the test particles.


\subsection{Numerical results with tsMC method}

Next, we show the numerical results of the HW entropy in qYM 
in the tsMC method
using the formula Eq.~\eqref{Eq:HWEtsMC}
.

\begin{figure}[t]
\begin{center}
\includegraphics[width=8cm]{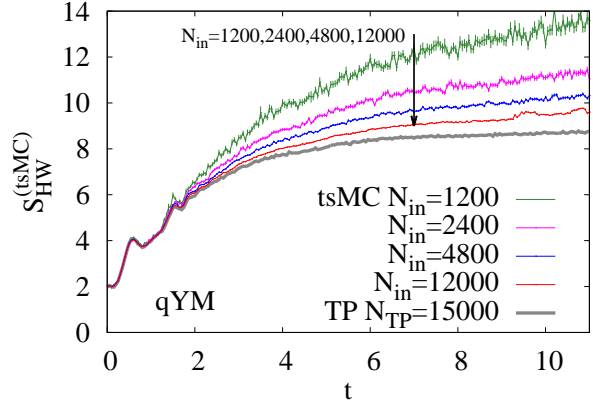}
\caption{Time dependence of HW entropy calculated by using tsMC method.}
\label{MCPA_QYMm}
\end{center}
\end{figure}


Figure \ref{MCPA_QYMm} shows the time evolution of the HW entropy
calculated 
in the tsMC method
with 
the sample numbers
$N_{\rm in}=1200, 2400, 4800$ and $12000$.
$N_{\rm out}$ is taken to be the same as $N_{\rm in}$.
The errors 
attached to $S_{\rm HW}$ in the 
present figure is estimated only for the Monte-Carlo integrals
outside of $\log$ in Eq.~\eqref{Eq:HWEtsMC},
and those from the integral inside the $\log$ is not taken into account, 
which causes an additional systematic error.

We see that the larger the value of $N_{\rm in}$, the smaller the HW entropy,
which is an opposite dependence on the sample number to that in the TP method. 
Nevertheless the gross behavior in the time evolution of the HW entropy 
is quite similar in the two methods apart from the tiny fluctuations; 
After showing an oscillatory behavior in a
first short period, it increases in a monotonous way and
its growth rate decreases gradually.
More quantitative comparison of the two methods
will be presented in the next subsection.

\subsection{Comparison of the two methods}
\label{subsec:CompTP}

\begin{figure}[t]
\begin{center}
\includegraphics[width=8cm]{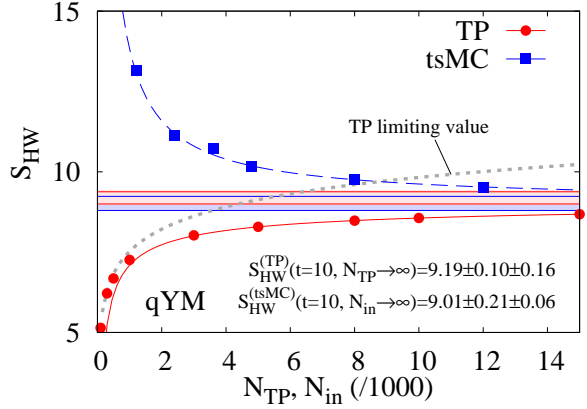}
\caption{HW entropy in qYM at $t=10$ as a function of $\NTP$ ($\NMC$),
and its extrapolation to infinitely large $\NTP$ ($\NMC$)
in the TP (tsMC) method.
Filled circles (squares) show TP (tsMC) results,
and the solid (dashed) line shows a fit function to TP (tsMC) results.
The dotted line is limiting value given by Eq.~\eqref{eq:limiting}.
The shaded areas show the extrapolated value 
in the limit of $\NTP, \NMC \to \infty$.}
\label{QYMmt10}
\end{center}
\end{figure}

Figure~\ref{QYMmt10} shows the HW entropy at $t=10$
as a function of $N_{\rm TP}$ ($N_{\rm in}$)
in the TP (tsMC) method. 
We fit a linear function $f(t)=at+b$
to the calculated $S_\mathrm{HW}(t)$ data
in the range $10-\Delta t\leq t \leq 10+\Delta t\ (\Delta t=1)$,
and adopt $f(t=10)$ as the HW entropy value at $t=10$.
This procedure provides a smoother curve 
and reduces the errors coming from fluctuations
compared to directly using the raw data.

The HW entropy in the TP method
becomes larger with increasing $N_{\rm TP}$ as already mentioned;
At $t=10$, 
    $S_{\rm HW}\simeq 5.1$ for $N_{\rm TP}=100$
and $S_{\rm HW}\simeq 8.7$ for $N_{\rm TP}=15000$. 
%
We also show 
the fit results to the 
data for larger samples, say $N_{TP}\ge 5000$, with a fit function,
\begin{align}
f(N)=a-\frac{b}{N^c}.
\label{eq:fitfunc}
\end{align} 
The extrapolated value to $N_{\rm TP}\rightarrow \infty$ is 
$9.19\pm 0.10$.  
When we use other fit functions such as
$f(N)=a-b/(N/c+1)$ and 
$f(N)=a-b/N+c/N^2$,
the fit results have differences with a standard deviation of $0.16$,
which should be considered as a systematic error.
Thus the HW entropy in the TP method is obtained as
\begin{align}
S_\mathrm{HW}^\mathrm{(TP)}(t=10)
=9.19 \pm 0.10~\mathrm{(stat.)} \pm 0.16~\mathrm{(syst.)}
\ .
\end{align}

With increasing $N_{\rm in}$,
the HW entropy calculated in the tsMC method decreases,
which is an opposite behavior to that in the TP method as noted before.
At $t=10$, 
$S_{\rm HW}\simeq 13.2$ for $N_{\rm in}=1200$
and 
$S_{\rm HW}\simeq 9.5$ for $N_{\rm in}=12000$.
%
We also show the fit results to the data.
We adopt Eq.~\eqref{eq:fitfunc} for the fit function.
%
From the fit results, the HW entropy in the tsMC method 
is found to be
\begin{align}
S_\mathrm{HW}^\mathrm{(tsMC)}(t=10)
=9.01 \pm 0.21~\mathrm{(stat.)} \pm 0.06~\mathrm{(syst.)}
\ ,
\end{align}
where the central value and the statistical error are obtained
from the fit using Eq.~\eqref{eq:fitfunc},
and the systematic error is evaluated from the fits using several fit functions
as done in the TP method.

\subsection{Discussions}

The time evolution of the HW entropies obtained in the TP and tsMC methods
shows a similar behavior with each other:
The HW entropy increases with an oscillatory behavior in the early stage,
then shows a monotonous increase with 
a decreasing rate.
The HW entropy at each $t$ in the TP method increases along with $\NTP$,
while it decreases with increasing $\NIN$ in the tsMC method.
Thus we can guess that the real value of the HW entropy lies between
the results in the TP and tsMC methods.
Actually, the extrapolated values at $t=10$,
$S_\mathrm{HW}^\mathrm{(TP)}(t=10)=9.19\pm0.10\pm0.16$ at $\NTP\to\infty$
and $S_\mathrm{HW}^\mathrm{(tsMC)}(t=10)=9.01\pm0.21\pm0.06$ at $\NIN\to\infty$ 
in the TP and tsMC methods respectively,
are consistent with each other within the error.
These results are also
in agreement with that in Ref.\cite{TsaiMuller}.

These two methods, TP and tsMC methods,
give consistent results after $N\rightarrow\infty$ extrapolation.
On the other hand, with finite number of $\NTP$ and $N_{\rm in}$,
they could give seemingly inconsistent results
depending on the dynamics.
We here have a deeper look at this issue.
In the tsMC method, the entropy seems to
keep increasing even for the later time, in contrast to 
the results in the TP method with finite $\NTP$
and in Ref.\cite{TsaiMuller}.
The discrepancy may come from 
the special shape of the potential:
there are two flat directions in the potential for the qYM system,
although the width of them tends to shrink at large distances.
Then, the classical trajectory can keep growing along the flat direction,
which would cause an unlimited spreading of the Husimi function
and a permanent increase
of the HW entropy calculated in the semiclassical approximation.
(In the case of the TP methods,
there exists limiting value of the HW entropy depending on $\NTP$,
which gives rise to the apparent saturation of $S$ at large $t$.
)
By comparison,
it is shown that the exact energy spectra of the qYM are all discrete ones,
because of the 
shrinking width leading to an increase of the kinetic energy due to 
the uncertainty relation,
 although the volume of $\{(p,q)|H(p,q)\le E\}$ 
is infinite \cite{Simon1983}.
Note that the discrete spectra implies that the wave functions of 
the energy eigen states are all bound.
Thus the corresponding Husimi function would not have a support 
at the infinite distance due to the quantum effect,
and the HW entropy may not show the ever increasing behavior
but have a saturated value.
This plausible conjecture can only be confirmed by a full quantum
calculation beyond the semiclassical approximation. 
Such a calculation is beyond the scope of the present work and
will be left as a future work.
Instead, we shall take another model, 
which is a modified version of the qYM one 
free from flat directions in its potential.

\section{Modified quantum Yang-Mills model}

\begin{figure}[t]
\begin{center}
\includegraphics[width=8cm]{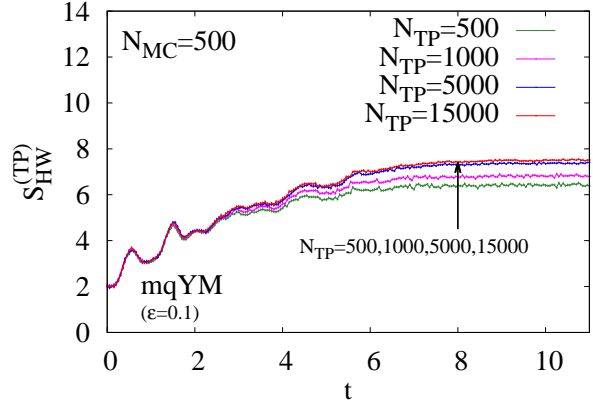}
\caption{Time dependence of HW entropy by using TP method in modified qYM.}
\label{TP_MQYMm}
\end{center}
\end{figure}

\begin{figure}[t]
\begin{center}
\includegraphics[width=8cm]{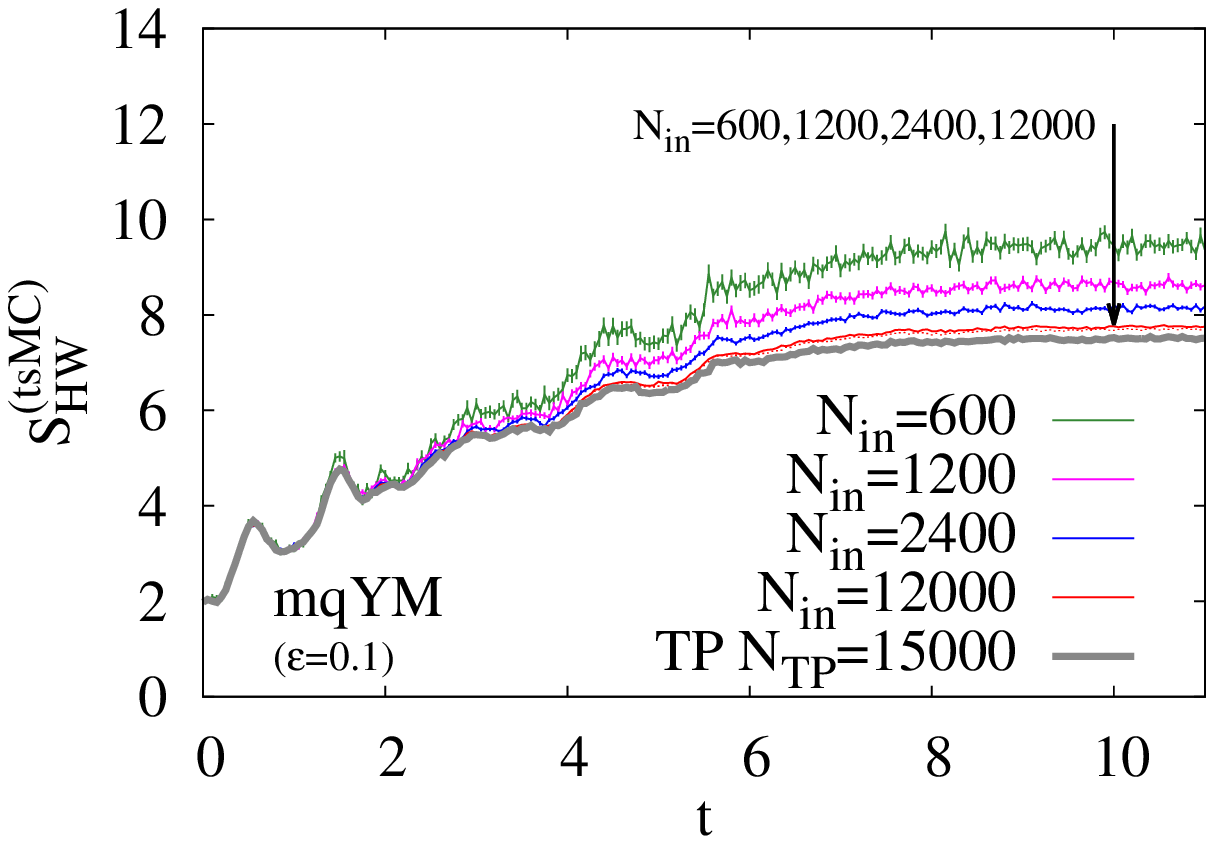}
\caption{Time dependence of HW entropy by using tsMC method in modified qYM.}
\label{MCPA_MQYMm}
\end{center}
\end{figure}

\begin{figure}[t]
\begin{center}
\includegraphics[width=8cm]{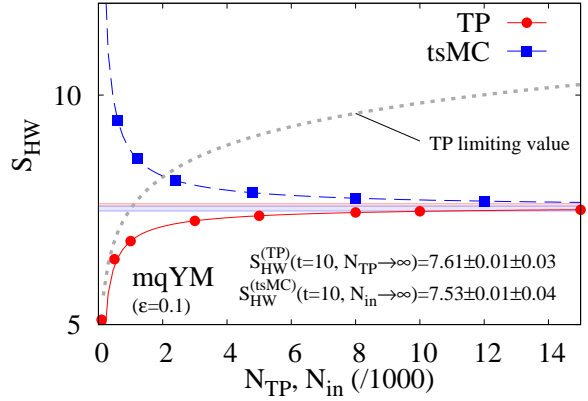}
\caption{HW entropy in mqYM at $t=10$ as a function of $\NTP$ ($\NMC$),
and its extrapolation to infinitely large $\NTP$ ($\NMC$)
in the TP (tsMC) method.
Filled circles (squares) show TP (tsMC) results,
and the solid (dashed) line shows a fit function to TP (tsMC) results.
The dotted line is limiting value given by Eq.~\eqref{eq:limiting}.
The shaded areas show the extrapolated value 
in the limit of $\NTP, \NMC \to \infty$.}
\label{mQYMmt10}
\end{center}
\end{figure}

Let us consider the model in which 
quartic potential terms are added to the qYM Hamiltonian; 
\begin{align}
H=\frac{p_1^2}{2m}+\frac{p_1^2}{2m}+\frac{1}{2}g^2q_1^2q_2^2+\frac{\epsilon}{4}q_1^4+\frac{\epsilon}{4}q_2^4.
\label{Eq:H_mqYM}
\end{align}
We call the system ``modified quantum Yang-Mills (mqYM)''. 
The system is studied in Ref.\cite{Sugita-Aiba2002,Sugita03}
with $g^2<0$ in the context of chaos.
It is apparent that 
there is no flat direction in the potential due to the quartic terms. 
We take $g^2=1$ and $\epsilon=0.1$ in the Hamiltonian,
Eq.~\eqref{Eq:H_mqYM}.
The mqYM system is found to be integrable
with $\epsilon/g^2=1, 1/3$ and $\infty$~\cite{Baker:1995bp,Sugita-Aiba2002}.
Our choice of $\epsilon/g^2=0.1$ is well apart from the integrable region.
Since $\epsilon$ is not very large,
the HW entropy shows a similar behavior to that in qYM at early times,
as shown later.

In this section,
we shall calculate the HW entropy of the mqYM system in
the TP and tsMC methods
.
The analyses are carried out in a similar way to those
for the qYM system.

In Figs.~\ref{TP_MQYMm} and \ref{MCPA_MQYMm},
we show the time evolution of the HW entropy in mqYM
calculated using the TP
($\NTP=500, 1000, 5000$ and $15000$ with $\NMC=500$)
and tsMC ($\NIN=600, 1200, 2400$ and $12000$) methods, respectively.
$N_{\rm out}$ is taken to be the same as $\NIN$ for tsMC.

The distribution function in Eq.~\eqref{Eq:initcond}
is used as the initial condition,
and the statistical errors are estimated for $\NMC$ ($\NIN$) samples
from a standard deviation in the TP (tsMC) method,
as in the qYM cases.

Both of the calculated results show that
the HW entropy first increases with an oscillatory behavior
and tends to saturate at later times, $t \gtrsim 6$.
%
%
%
The later-time $\SHW$ values depend on the sample number,
$\NTP$ and $\NIN$;
With increasing $\NTP$ ($\NIN$), the HW entropy increases (decreases)
in the TP (tsMC) method.
These are the features also found in qYM. By comparison, it
should be noted
that there seems to be saturation 
of $\SHW$ both in the TP and tsMC methods in mqYM,
in contrast to qYM.
This may be originated from the 
finite phase space volume where the Husimi function has a support.

In Fig.~\ref{mQYMmt10},
we show the HW entropy at $t=10$ as a function of $\NTP$ or $\NIN$.
We fit a linear 
function
to calculated $S_\mathrm{HW}(t)$ results in the range $9<t<11$,
and adopt 
$f(t=10)$ as the HW entropy value
at $t=10$.
In the TP method, $S_\mathrm{HW}^\mathrm{(TP)}(t=10)\simeq 6.4$ and $7.5$
for $\NTP=500$ and $15000$, respectively.
In tsMC, we find
$S_{\rm HW}^\mathrm{(tsMC)}(t=10)\simeq 9.4$
and $7.7$ for $\NIN=600$ and $12000$, respectively.

%
%
%

The extrapolated values of $\SHW$ at 
$\NTP\to\infty$ and $\NIN\to\infty$
are found to be
\begin{align}
S_\mathrm{HW}^\mathrm{(TP)}(t=10)
=7.61\pm0.01\mathrm{(stat.)}\pm0.03\mathrm{(syst.)}\ ,
\\
S_\mathrm{HW}^\mathrm{(tsMC)}(t=10)
=7.53\pm0.01\mathrm{(stat.)}\pm0.04\mathrm{(syst.)}\ ,
\end{align}
in the TP and tsMC methods, respectively.
The central values and the statistical errors are obtained
from the fit using Eq.~\eqref{eq:fitfunc},
and the systematic error is evaluated from the fits using several fit functions.
These two values are consistent with each other within the error.

The observation shows that the two methods, tsMC and TP, 
are especially effective for such a potential which bounds Husimi function
in finite region.
Thus, we are confident of the validity of the two methods
in the mqYM system. 


\section{Summary}

We have discussed entropy creation in isolated quantum systems
by using the Husimi-Wehrl entropy evaluated 
in a semiclassical treatment.
The semiclassical treatment is known to be useful in some of the systems
such as the
inflation in early universe 
and the early stage of relativistic heavy ion collisions.
These systems are expected to bear instabilities and/or chaoticities
in their classical counter systems,
then the smearing of the phase space distribution by the minimal wave packet
causes the entropy production in terms of the Wehrl entropy or the $H$ function
of Boltzmann even in isolated quantum systems.
This is nothing but the Husimi-Wehrl entropy, the Wehrl entropy obtained 
by using the Gaussian smeared Wigner function (Husimi function)
for the phase space distribution.

The semiclassical time evolution of the Husimi function is given 
by solving a classical equation of motion and smearing with 
a Gaussian packet.
Combining this semiclassical treatment
with the Monte-Carlo numerical integral technique,
we have developed two methods,
the test-particle (TP) method and the two-step Monte Carlo (tsMC) method.
We have
applied these two methods
to quantum mechanical systems in two dimensions, 
the quantum Yang-Mills (qYM) and the modified quantum Yang-Mills (mqYM)
systems.
The classical counter systems of these are known to be chaotic.
We have demonstrated that the Husimi-Wehrl entropy obtained in the TP (tsMC)
method approaches the converged value from below (from above) 
with an increasing sample number, then
we can 
guess the true value of HW entropy.
We have further found that the results 
of the TP and tsMC methods in the infinite sampling number limit
are consistent within the error.
Therefore, the simultaneous application of the two methods ensures
the reliability of the results of the Husimi-Wehrl entropy at a given time.

The extension of our methods to a multidimensional system 
is 
straightforward.
We expect that these methods are
useful in systems with many degrees of freedom 
such as the quantum field theory.
These methods are, in principle, applicable to higher-dimensional problems,
and we have confirmed that they actually work in three and four dimensional
systems. In higher dimensions, we need much more Monte-Carlo samples to
obtain
statistically reliable results, and it would be necessary to make some
approximations
for practical purposes.
Work in this direction is in progress.

\section*{Acknowledgement}
We would like to thank Ayumu Sugita for a good lecture and useful suggestions.
%
This work was supported in part by 
the Grants-in-Aid for Scientific Research from JSPS
 (Nos.
 20540265, 
 23340067, 
 24340054, 
 24540271, 
 15K05079
),
the Grants-in-Aid for Scientific Research on Innovative Areas from MEXT
 (Nos. 23105713, 
       24105001, 24105008 
),
and
 by the Yukawa International Program for Quark-Hadron Sciences.
T.K. is supported by the Core Stage Back Up program in Kyoto University.

\appendix

%

\section{HW entropy in inverted harmonics}
The inverted harmonic oscillator (IHO) is an unstable system,
where the Hamiltonian is given as
\begin{equation}
H=\frac{p^2}{2 m}-\frac{1}{2}\lambda^2 q^2.
\end{equation}
In this system, the classical trajectories are not restricted
in a finite region, but extends to infinitely large spatial and momentum regions.
While this unbounded nature makes the numerical calculation difficult,
the analytic expression of the HW entropy is known~\cite{KMOS}.
Then by comparing the numerical results with the analytic solution,
we can examine the validity and the precision of the numerical methods.

\subsection{Analytic solution}
When the initial distribution of Wigner function is given by a Gaussian,
\begin{equation}
f_W(p,q;t=0)
=2 \exp\left(-\frac{1}{\hbar \omega}p^2 - \frac{\omega}{\hbar}q^2\right),
\end{equation}
the time evolution of the HW entropy is obtained analytically \cite{KMOS}. 
Since the potential is quadratic, the semiclassical analysis is exact, 
and the time evolution of the Wigner function is calculated
by solving the classical equation of motion. 
The HW entropy at time $t$ is given as~\cite{KMOS}, 
\begin{align}
&S_{HW}(t)=\log\frac{\sqrt{A(t)}}{2}+1\ ,
\\
&A(t)=2(\sigma \rho \cosh(2\lambda t) +1+\delta \delta')\ ,
\\
&\sigma=\frac{\lambda^2+\omega^2}{2 \lambda \omega}\ ,
\quad
\delta=\frac{\lambda^2-\omega^2}{2 \lambda \omega}\ ,
\\
&\rho=\frac{\Delta^2+\lambda^2}{2 \Delta \lambda}\ ,
\quad
\delta'=\frac{\Delta^2-\lambda^2}{2 \Delta \lambda}.
\end{align}

\begin{figure}[htb]
\begin{center}
\includegraphics[width=8cm]{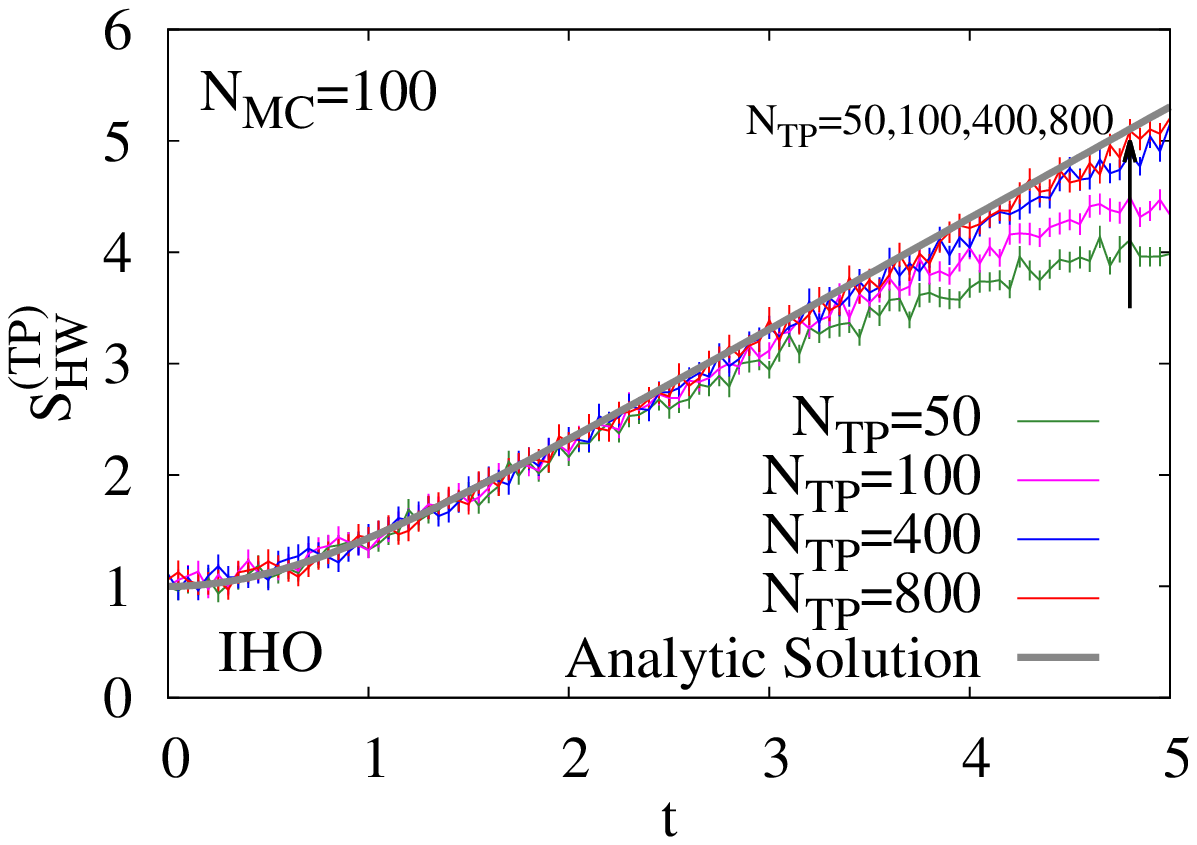}
\caption{Time dependence of the HW entropy of the inverted harmonic oscillator
in the TP method.}
\label{TP_inv}
\end{center}
\end{figure}

\subsection{Numerical results with TP}


Figure \ref{TP_inv} shows the time evolution of the HW entropy of IHO
calculated in the TP method with $\NTP=50-800$.
We find that the TP method can well describe the time evolution 
of the HW entropy at early times,
and that numerical results show saturated behavior in later times.
Since there exists a limiting value of $\SHW$ in the TP method
as discussed in Subsec.~\ref{subsec:qYM-TP},
we need to take a large number of $\NTP$
to describe a large amount of entropy production.
It should be noted that numerical results converge 
in the limit of $\NTP\to\infty$,
and the converged result well describe the analytic result.

\begin{figure}[htb]
\begin{center}
\includegraphics[width=8cm]{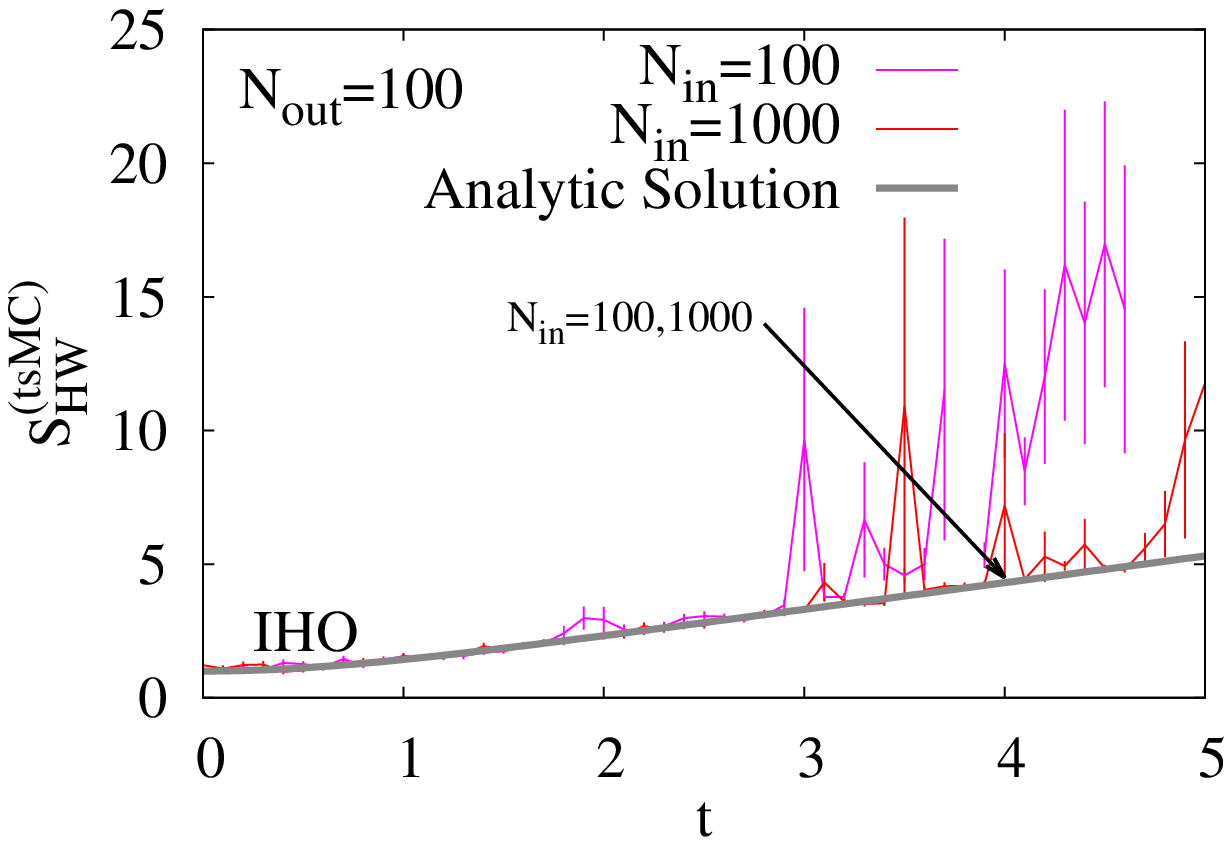}
\caption{Time dependence of the HW entropy of the inverted harmonic oscillator
in the tsMC method.}
\label{MC_inv}
\end{center}
\end{figure}

\subsection{Numerical results with tsMC}

Figure \ref{MC_inv} shows the time evolution of the HW entropy of IHO
in the tsMC method with $\NIN=N_\mathrm{out}=100$ and $1000$.
We find that numerical results are consistent with the analytic solution
at early times $t\le 3$, 
but that the numerical results tend to overestimate the analytic results
and numerical errors become very large at later times.
The large error would come from the poor overlap between
the Wigner function and the coarse-graining Gaussian function at later time,
which makes importance sampling spoiled.
On the other hand, the systematic overestimation may be due to the lack of
sampling points in the Monte-Carlo integration in the logarithmic
function (See Eq.~\eqref{Eq:HWEtsMC}).
We note here that the statistical-error estimation is performed only for
the Monte-Carlo integration outside the log.



\end{document}